\documentstyle[aps,floats,prd,psfig]{revtex} 
     
\newcommand{\dsz}{{\rm Rei}}

\newcommand{\Mhz}{\mbox{ MHz}}

\newcommand{\kel}{\mbox{ K}}

\newcommand{\ApJ}{Astrophys. J.}

\newcommand{\MNRAS}{Mon. Not. Roy. Astron. Soc.}

\newcommand{\da}{d_A}

\newcommand{\lin}{{}}
\newcommand{\bn}{\hat{\bf n}}    
    
\newcommand{\xh}{{x_H}}

\newcommand{\veck}{{\bf k}}

\newlength{\tskip}
\newlength{\colwidth}

\newcommand{\beq}{\begin{equation}}
\newcommand{\eeq}{\end{equation}}
\newcommand{\beqa}{\begin{eqnarray}}
\newcommand{\eeqa}{\end{eqnarray}}

\long\def\comment#1{}

\newcommand{\wjm}{\left(
                           \begin{array}{ccc}
         l_1 & l_2  & l_3  \\
         m_1 & m_2  & m_3
                           \end{array}
                   \right)}

\newcommand{\bi}{B_{l_1 l_2 l_3}}

\newcommand{\deld}{\delta^{\rm D}}
\newcommand{\bm}{\hat{\bf m}}
\newcommand{\bl}{\hat{\bf l}}
\newcommand{\bk}{{\bf k}}
\newcommand{\rad}{r}    
\newcommand{\dop}{{\rm dop}}

\newcommand{\Ylm}[1]{Y_{l_#1}^{m_#1}}

\newcommand{\Ylmn}{Y_{l}^{m}}

\newcommand{\alm}[1]{a_{l_#1 m_#1}}

\begin{document}   

\twocolumn[\hsize\textwidth\columnwidth\hsize\csname @twocolumnfalse\endcsname
\title{Cross-Correlation Studies between CMB Temperature Anisotropies and 21 cm Fluctuations}
\author{Asantha Cooray}
\address{California Institute of Technology, Mail Code
130-33, Pasadena, CA 91125\\
E-mail: asante@tapir.caltech.edu}

\maketitle
\begin{abstract}
During the transition from a neutral to a fully reionized universe, scattering of 
cosmic microwave background (CMB) photons via free-electrons leads to a new anisotropy contribution to the temperature distribution.
If the reionization process is inhomogeneous and patchy, the era of reionization is also visible via
brightness temperature fluctuations in the redshifted 21 cm line emission
from neutral Hydrogen. Since regions containing electrons and neutral Hydrogen are
expected to trace the same underlying density field, the two are (anti) correlated and this is expected to be
reflected in the anisotropy maps via a correlation between arcminute-scale CMB  temperature and the
21 cm background. In terms of the angular cross-power spectrum, unfortunately,
this correlation is insignificant due to a geometric cancellation associated with second order CMB anisotropies.
The same cross-correlation between ionized and neutral regions, however, can  be studied using a bispectrum 
involving large scale velocity field of ionized regions from the Doppler effect,
arcminute scale CMB anisotropies during reionization, and the 21 cm background. 
While the geometric cancellation is partly avoided, the signal-to-noise ratio related to this bispectrum is reduced due to the large
cosmic variance related to velocity fluctuations traced by the Doppler effect. 
Unless the velocity field during reionization can be independently established, 
it is unlikely that the correlation information related to the relative distribution of ionized electrons and regions containing 
neutral Hydrogen can be obtained with a combined study involving CMB and 21 cm fluctuations.
\end{abstract}

\hfill
]

\section{Introduction}

The large angle bump in the polarization-temperature cross correlation power spectrum 
\cite{Zaldarriaga:1996ke} measured in WMAP data \cite{Bennett:2003bz}
indicates an optical depth to electron scattering of 0.17 $\pm$ 0.04 \cite{Kogetal03}.
To explain both this high optical depth and the Lyman-$\alpha$ optical depth from Gunn-Peterson
troughs \cite{GunPet65} towards $z \sim 6$ quasars in the Sloan Digital Sky Survey \cite{Fan:2001vx} requires
a complex reionization history \cite{Cen:2003ey,Chen:2003sw,Cen:2002zc}.
While slight modifications to the large angular scale polarization
power spectra exist with different reionization histories that integrate to the same electron scattering optical depth,
one cannot use these changes to fully reconstruct the reionization history as a function of redshift 
due to large cosmic variance associated with anisotropy measurements at a few tens degree angular scales \cite{Kaplinghat:2002vt}.

Under standard expectations for reionization, mainly due to UV light emitted by first luminous objects,
the reionization process is expected to be both patchy and inhomogeneous \cite{Barkana:2000fd}. This leads to fluctuations in the
electron scattering optical depth and to a modulation of both temperature and polarization contributions such that new anisotropy fluctuations
are generated at arcminute scales corresponding to inhomogeneities in the visibility function \cite{Hu:1999vq,Santos:2003jb}. 
The increase in sensitivity and angular resolution of upcoming CMB anisotropy data suggests that
such small-scale fluctuations can soon be used to understand the reionization history and associated physics beyond measurements related to
large-scale polarization alone. 

The secondary reionization-related anisotropies, unfortunately, are expected to be 
confused with other secondary effects such as the Sunyaev-Zel'dovich (SZ; \cite{SunZel80}) contribution 
at late times associated with galaxy clusters and first supernovae \cite{Oh:2003sa},
 and higher order effects such as gravitational lensing \cite{Hu:ee}. Though this confusion can be partly removed,
such as through frequency cleaning in the case of SZ \cite{Cooray:2000xh} or subtraction of lensing via higher order
statistics \cite{Okamoto:2003zw}, secondary anisotropies alone cannot be used to
extract the detailed history of reionization beyond the integrated optical depth \cite{Santos:2003jb,Zhang:2003nr}. 

An interesting possibility involves
the cross-correlation between small-scale CMB polarization maps and images of the high redshift universe. This cross-correlation
can help distinguish broad aspects such as whether the universe reionized once or twice \cite{Cooray:2003dt}.
Beyond the reionization history, mainly in the scattering visibility function as a function of redshift,
it would be interesting to study additional details related to the reionization process such as the
relative distribution of free-electrons and neutral Hydrogen. Since 
the brightness temperature fluctuations associated with the 21 cm spin-flip transition \cite{Fie58}
trace the neutral Hydrogen distribution, 
the combination involving secondary CMB and 21 cm fluctuations \cite{Tozzi:1999zh} could then provide additional details related to
reionization.

In this paper, we consider such a combined study in the form of a  cross-correlation between arcminute scale CMB anisotropy maps and
images of the high redshift universe around the era of reionization from the 21 cm background. We make the
assumption that the reionization process is inhomogeneous and patchy such that at certain epochs during partial ionization, contributions
are generated to both the 21 cm background  and CMB.   In such an era, we
also assume that the spin temperature of the neutral Hydrogen distribution is decoupled from CMB and
is dominated by the kinetic temperature; In this case, the signature in 21 cm emission is fluctuations related to the distribution of
neutral Hydrogen, though, in the limit where the spin temperature is smaller than that of CMB (prior to reionization and appearance of first sources), 
21 cm fluctuations will be tightly coupled to that of large angular scale CMB temperature; In this case, one naturally expects 
a perfect cross-correlation between these two maps. We do not consider this possibility as the reionization process
is expected to heat the IGM by $z \sim 20$. Also, opportunities for very low frequency  observations,
where 21 cm signatures from redshifts prior to reionization are expected, are extremely limited.

In the era of partial reionization, the existence of the proposed cross-correlation is due to the fact that
inhomogeneities that lead to fluctuations in both CMB (in terms of the electron distribution) and 21 cm background (via the neutral
content) trace the same underlying density field. One expects the two to
spatially correlate though this would be an anti-correlation as regions containing free-electrons will be mostly free of
neutral material. This fact is captured by a spatial cross-power spectrum between free-electron and neutral
Hydrogen fluctuations. Since the 21 cm background allows one to probe the power spectrum of neutral Hydrogen alone,
while CMB probes the power spectrum of reionized patches, in combination, the cross-power spectrum provides
additional information on physics related to reionization.

Unfortunately, while there is a strong (anti) cross-correlation between  free-electrons and neutral
Hydrogen fluctuations,  we find the observable projected angular cross-spectrum
to be insignificant due to a geometric cancellation. This cancellation comes from
the  CMB side  and involves the line of sight projection of the velocity field during reionization.
The same cross-correlation can also be studied using a higher order correlation in the form of a bispectrum in
Fourier space involving the large scale velocity field of ionized regions from the Doppler effect,
arcminute scale CMB anisotropies, and the 21 cm background.  This measurement avoids the geometric cancellation associated with the
line of sight projection, but due to the large cosmic variance associated with the velocity field traced by the Doppler
effect, its measurement is limited to signal-to-noise ratios of order ten (with noise dominated maps) or, at most, a hundred.

We discuss the measurement of proposed cross-correlations 
using CMB maps from upcoming missions, such as Planck surveyor and the South Pole Telescope (SPT), 
and maps of the 21cm background with, say, the 
Square Kilometer Array. Unfortunately, 
with signal-to-noise ratios around ten or below, it is unlikely that one can use the proposed
cross-correlation bispectrum to easily extract detailed information on the relative distribution between neutral Hydrogen and electrons.

The paper is organized as follows.  In \S~\ref{sec:deriv}, we derive the existence of the cross-correlation
both in terms of the cross power spectrum between 
CMB temperature and 21 cm fluctuations and a higher order bispectrum that avoids a geometric cancellation associated with the
cross power spectrum. In \S~\ref{sec:results}, we discuss our results and suggest that though there is adequate 
signal-to-noise to perform a cross correlation study in terms of the bispectrum. We briefly discuss how this measurement
can be improved and what information related to reionization can be extracted from this measurement.
We conclude with a summary in \S~\ref{sec:summary}.

\section{Calculation Method}
\label{sec:deriv}

We first discuss fluctuations in the 21 cm background and then small-scale CMB anisotropies related to reionization. We will then
consider the relation between these two quantities in terms of the angular power spectrum related to the cross-correlation. 
Beyond the power spectrum,
we also discuss a bispectrum associated with these two quantities and the large angular scale CMB fluctuations related to 
velocity variations during reionization.

\subsection{21 cm Fluctuations}

In the low optical depth limit of the radiation transfer,
the  brightness temperature fluctuation associated with 21 cm background can be written
as \cite{Zaldarriaga:2003du}
\beqa
\delta T_{\rm 21 cm}(\nu) & \approx & \frac{T_S - T_{\rm CMB}}{1+z} \, \tau_{H}
\label{eq:dtb} 
\eeqa
where $T_{\rm CMB} = 2.73 (1+z) \kel$ is the CMB temperature at
redshift $z$, $T_S$ is the spin temperature of the IGM, 
and the optical depth in the 21 cm hyperfine transition, $\tau_{H}$, is
\beqa
\tau_H & = & \frac{ 3 c^3 \hbar A_{10} \, n_{\rm HI}}{16
k \nu_0^2 \, T_S \, H(z) }
\label{eq:tauigm} \\
\, & \approx & 2.7 \times 10^{-3}  \left( \frac{\Omega_b h^2}{0.02} \right) (1+\delta_g) \xh \left[
\frac{T_{\rm CMB}(z)}{T_S} \right] \sqrt{1+z} \, .
\nonumber
\eeqa
Here $n_{\rm HI}$ is the neutral hydrogen density expressed in terms of fluctuations in the density as
$n_{\rm HI}=x_H\bar{n}_g(1+\delta_g)$, where $\bar{n}_g$ is the mean number density of cosmic baryons and
$x_H$ is the ionization fraction, $\nu_0=1420.4 \Mhz$ is the rest-frame hyperfine transition
frequency, $A_{10}$ is the spontaneous emission coefficient for the transition ($2.85 \times 10^{-15}$ s$^{-1}$).
We refer the reader to Ref.~\cite{Zaldarriaga:2003du} for further details.

We expand fluctuations in the brightness temperature related to the 21 cm emission and write multipole moments
as
\begin{equation}
a_{lm}^{\rm 21 cm} = \int d\bn \delta T_{\rm 21 cm}(\bn) \Ylmn {}^*(\bn) \, ,
\end{equation}
with
\begin{eqnarray}
&&\delta T_{\rm 21 cm}(\bn) = \nonumber \\
&& {23.2 \; \rm mk} \left( \frac{\Omega_b h^2}{0.02} \right) [1+\delta_g(\bn)] \bar{\xh} \left[
\frac{T_s-T_{\rm CMB}}{T_S} \right] \sqrt{1+z} \, ,
\end{eqnarray}
where $\bar{\xh}$ is the mean neutral fraction.
Making use of the Rayleigh expansion
\begin{equation}
e^{i{\bf k}\cdot \hat{\bf n}\rad}=
4\pi\sum_{lm}i^lj_l(k\rad)Y_l^{m \ast}(\hat{\bk})
\Ylmn(\bn)\,,
\label{eqn:Rayleigh}
\end{equation}
we can now write the multipole moments as
\begin{eqnarray}
a^{\rm 21cm}_{lm} &=& 4 \pi i^l \int \frac{d^3\veck}{(2 \pi)^3}
\delta(\veck)  I_l^{\rm 21 cm}(k) \Ylmn(\hat{\veck}) \, , \nonumber\\
I_l^{\rm 21cm}(k) &=& \int d\rad  W^{\rm 21cm}(k,\rad)j_{l}(k\rad) \, ,
\label{eqn:secondaryform}
\end{eqnarray}
where the window function is
\begin{equation}
W^{\rm 21cm} = 23.2\; {\rm mk}\;  \left( \frac{\Omega_b h^2}{0.02} \right) \xh \left[
\frac{T_s-T_{\rm CMB}}{T_S} \right] (1+z)^{1/2}  \, .
\end{equation}
In numerical calculations, we consider 21 cm measurements during partial reionization ($0 < x_H < 1$) 
where the spin temperature is expected to be dominated by
the kinetic temperature. In this case $T_s \sim T_k \gg T_{\rm CMB}$ and the window function related to the 21 cm fluctuations is
independent of the spin temperature of neutral Hydrogen.

In Eq.~\ref{eqn:secondaryform}, $r(z)$ is the conformal distance (or lookback time) from the observer at redshift $z=0$, given by
\begin{equation}
\rad(z) = \int_0^z {dz' \over H(z')} \,,
\end{equation}
where the expansion rate for adiabatic CDM cosmological models with a cosmological constant is
\begin{equation}
H^2 = H_0^2 \left[ \Omega_m(1+z)^3 + \Omega_K (1+z)^2
              +\Omega_\Lambda \right]\,,
\end{equation}
where $H_0$ can be written as the inverse
Hubble distance today $H_0^{-1} = 2997.9h^{-1} $Mpc.
We follow the conventions that
in units of the critical density $3H_0^2/8\pi G$,
the contribution of each component is denoted $\Omega_i$,
$i=c$ for the CDM, $b$ for the baryons, $\Lambda$ for the cosmological
constant. We also define the
auxiliary quantities $\Omega_m=\Omega_c+\Omega_b$ and
$\Omega_K=1-\sum_i \Omega_i$, which represent the matter density and
the contribution of spatial curvature to the expansion rate
respectively. Although we maintain generality in all derivations, we
illustrate our results with the currently favored $\Lambda$CDM
cosmological model. The parameters for this model
are $\Omega_c=0.30$, $\Omega_b=0.05$, $\Omega_\Lambda=0.65$ and $h=0.7$.

Using the multipole moments,
note that the angular power spectrum related to the 21 cm fluctuations is 
given by 
\begin{eqnarray}
\langle \alm{1}^* \alm{2}\rangle = \deld_{l_1 l_2} \deld_{m_1 m_2}
        C_{l_1}\,,
\end{eqnarray}
such that
\begin{equation}
C_l^{\rm 21 cm} = \frac{2}{\pi} \int k^2 dk P_{gg}(k)
                I_l^{\rm 21 cm}(k) I_l^{\rm 21 cm}(k) \,,
\label{eqn:clsecondary}
\end{equation}
where $P_{gg}(k)$ is the power spectrum of neutral Hydrogen distribution.
In Fig.~1(b), we show, as an example, the angular power spectrum of 21 cm fluctuations under the assumption that the
neutral Hydrogen fluctuations trace the linear density field with a bias factor given by the halo based calculations 
\cite{Cooray:2002di} such that $P_{gg}(k)=b_{x_H}^2 P^{\rm lin}(k)$ and 
the bias follows from
\begin{equation}
b_i(z) =  \frac{\int_{M_-}^{M_+} dM\, M\, \frac{dN}{dM} b(M,z)}{\int_{M_-}^{M_+} dM\, M\, \frac{dN}{dM}} \, .
\label{eqn:bias}
\end{equation}
Here, $M_-$ and $M_+$ are the lower and upper limits of masses, $dN/dM$ is the mass function \cite{Press:1973iz} 
and $b(M,z)$ is the halo bias of Ref.~\cite{Mo96}. For neutral Hydrogen fluctuations, we set
$M_- \rightarrow 0$ and $M_+$ the value corresponding to virial
temperatures of 10$^4$K. Note that we have taken a simple description of IGM bias here  and
the situation is likely to be more complicated. Since our main objective
is to see if there is an adequate signal-to-noise for a detection of the cross correlation between CMB and 21 cm fluctuations,
we ignore subtle details associated with neutral Hydrogen distribution.
In Fig.~1(b), for reference, we also show other contributions to the brightness temperature fluctuations at these
low frequencies, such as CMB and free-free  emission  from ionized halos \cite{Cooray:2004vd}.

For cross-correlation purposes with secondary order effects in CMB (discussed in the next Section), one needs to consider further fluctuations in the
21 cm background. In the patchy reionization scenario, these are naturally present during the epoch when the universe transits
from a neutral one to a fully reionized one. Making use of the inhomogeneities in the neutral gas distribution, we write
\beqa
\tau_H & \approx & 2.7 \times 10^{-3}  \left( \frac{\Omega_b h^2}{0.02} \right) (1+\delta_g) \nonumber \\
&\times& \xh (1+\delta_{x_H})\left[
\frac{T_{\rm CMB}(z)}{T_S} \right] (1+z)^{1/2} \, ,
\nonumber
\eeqa
in terms of fluctuations in the neutral Hydrogen fraction.
The brightness temperature fluctuations in the 21 cm background are
\begin{eqnarray}
&& \delta T_{\rm 21 cm}(\bn)= 
\int d\rad \int \frac{d^3{\bf k}_1}{(2\pi)^3} \int \frac{d^3{\bf
k}_2}{(2\pi)^3} \nonumber\\
&& \times W^{\rm 21cm}(\rad)
\delta_g({\bf k}-{\bf k'})\delta_{x_H}({\bf k'}) e^{i{\bf k}\cdot \hat{\bf n}\rad}\nonumber\\
\label{eqn:temp2}
\end{eqnarray}
with the same window function as before (and again independent of the spin temperature).
In Eq.~\ref{eqn:temp2}, the multiplication between the density fluctuations and ionization fraction inhomogeneities in real space
has been converted to a convolution between the two fields in Fourier space.
Setting, ${\bf k_1}={\bf k}-{\bf k'}$ and ${\bf k_2}={\bf k'}$,
the multipole moments can be simplified as
\begin{eqnarray}
&&a_{lm}^{\rm 21 cm} = 
\int d\rad
\int \frac{d^3{\bf k}_1}{(2\pi)^3} \int \frac{d^3{\bf
k}_2}{(2\pi)^3}
\sum_{l_1 m_1}\sum_{l_2 m_2}\sum_{m'} \nonumber\\
&& \times
(i)^{l_1+l_2} W^{\rm 21cm}(\rad)
j_{l_1}(k_1\rad) j_{l_2}(k_2\rad)
\delta_g({\bf k_1})\delta_{x_H}({\bf k_2})
\nonumber\\
&& \times
Y_{l_1}^{m_1}(\hat{\veck}_1) 
Y_{l_2}^{m_2}(\hat{\veck}_2)  M_{l l_1 l_2}^{m^\ast m_1^\ast
m_2^\ast}(\hat{\bf n}) \, ,
\label{eqn:alm21}
\end{eqnarray}
where we have defined a general integral over
spherical harmonics written such that
\begin{equation}
M_{l_1 l_2 ... l_i}^{m_1 m_2 ... m_i}(\hat{\bf n}) = \int d\hat{\bf n}
Y_{l_1}^{m_1}(\hat{\bf n}) Y_{l_2}^{m_2}(\hat{\bf n})
.... Y_{l_i}^{m_i}(\hat{\bf n}) \, .
\end{equation}

\subsection{CMB fluctuations}

The bulk flow of electrons that scatter the CMB photons, in the reionized epoch,
lead to temperature fluctuations through the well known Doppler effect \cite{Kai84}
\begin{equation}
T^\dop(\bn) = \int_0^{\rad_0} d\rad g(r) \bn \cdot {\bf v}_g(\rad,\bn
\rad)\, ,
\label{eqn:doppler}
\end{equation}
where ${\bf v}_g$ is the baryon velocity.
The visibility function, or the probability of scattering within $d\rad$ of $\rad$, is
\begin{equation}
g =  \dot \tau e^{-\tau} = X_e(z) H_0 \tau_H (1+z)^2 e^{-\tau}\,.
\end{equation}
Here
$\tau(r) = \int_0^{\rad} d\rad \dot\tau$ is the optical depth out to $r$,
$X_e(z)$ is the ionization fraction, as a function of redshift, and
\begin{equation}
       \tau_H = 0.0691 (1-Y_p)\Omega_b h\,,
\end{equation}
is the optical depth to Thomson
scattering to the Hubble distance today, assuming full
hydrogen ionization with primordial helium fraction of $Y_p (=0.24)$.

Here, we make use of a description for $X_e(z)$ that is consistent with current data such as the WMAP optical depth
\cite{Kogetal03}, following Ref.~\cite{Chen:2003sw}. The description is based on the 
Press-Schechter \cite{Press:1973iz} mass function and related to the reionization by UV light from
the first-star formation. Here, $X_e(z)$ varies from a value less than 10$^{-1}$ at a redshift of 30
to a value of unity when the universe is fully reionized, at a redshift of $\sim$ 5. The reionization history is rather broad
and the universe does not become fully reionized till a late time, though the reionization process began at a much higher epoch.
The total optical depth related to this ionization history is $\sim$ 0.15 consistent with WMAP analysis. 
The scattering is distributed widely given the long 
reionization process.  This is the same model as model A of Ref.~\cite{Cooray:2003dt}. Given the ionization fraction,
we set the neutral fraction, related to the 21 cm background, to be simply $x_H(z) = 1-X_e(z)$.

\begin{figure}[!h]
\centerline{\psfig{file=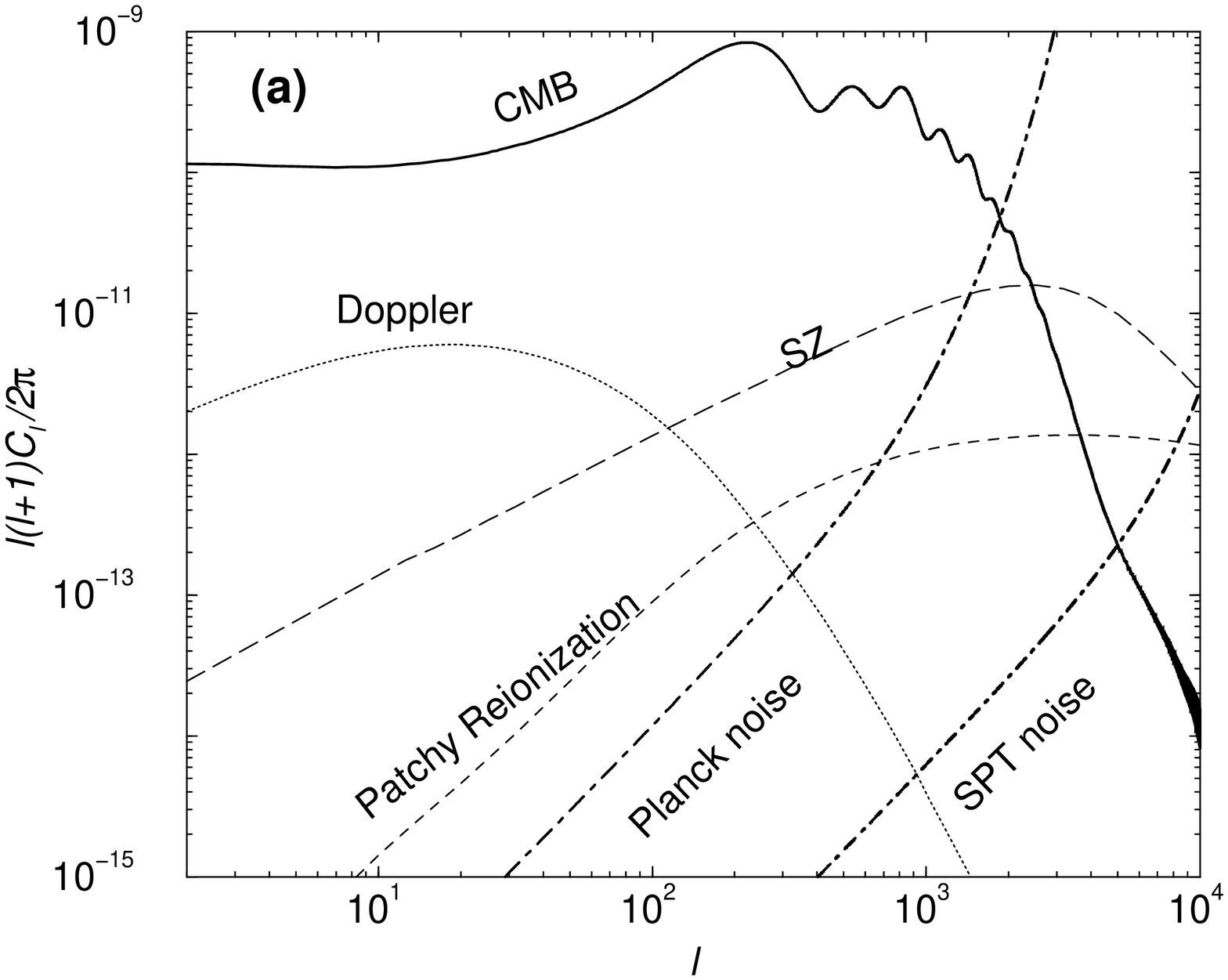,width=3.4in,angle=-90}}
\centerline{\psfig{file=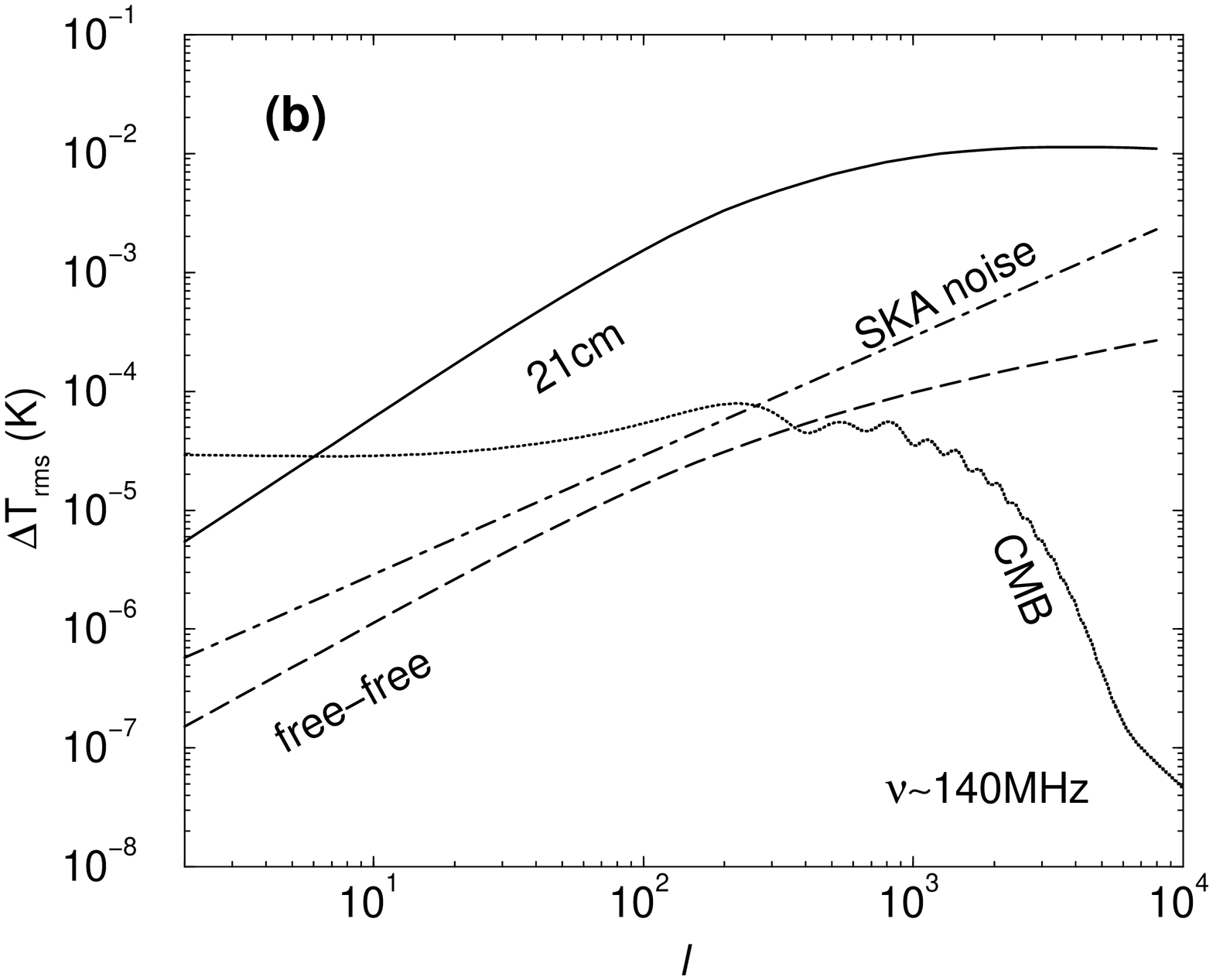,width=3.4in,angle=-90}}
\caption{(a) Power spectrum for the temperature
anisotropies in the fiducial $\Lambda$CDM model
with $\tau=0.15$. In addition to primary CMB fluctuations, we also show the
Doppler, SZ and Patchy Reionization contributions at late times during the reionization epoch
and afterward. For reference, we also show the
instrumental noise power spectra of Planck and the South Pole Telescope; We will use these for signal-to-noise calculations.
(b) The brightness temperature fluctuations at $\nu_{\rm
obs}=140$~MHz. Here, we show the 21 cm signal in a 1 MHz bandwidth window
around $z=10$, the CMB, and expected foreground contamination related to free-free fluctuations in 
ionized halos at redshifts around 3 and below.  For reference, here we also show the
instrumental noise contribution related to Square Kilometer Array (SKA) observations of the
21 cm fluctuations.}
\label{fig:cl}
\end{figure}

In Fig.~\ref{fig:cl}(a), we show the linear Doppler effect,
including contributions resulting due to double scattering effect
described in \cite{Kai84} (see, Ref.~\cite{Cooray:1999kg} for
details) following the reionization history. The power spectrum is such that it
peaks around the horizon at the reionization projected on the sky
today. As shown in Fig.~\ref{fig:cl}(a), the Doppler effect cancels out significantly at scales smaller than the
horizon at reionization surface since photons scatter against the crests and
troughs of the perturbation resulting in a net cancellation of the anisotropy.

The patchy reionization temperature fluctuations can be written as a product of the
line of sight velocity, under linear theory, and  fluctuations in the electron 
fraction, $x_e$ as
\begin{eqnarray}
T^\dsz(\hat{\bf n})&=&  \int d\rad
        g(r) \hat{\bf n} \cdot {\bf v}_g(r,\bn r) [1+\delta_{x_e}(r, \bn r)]
\nonumber\\
&=&-i \int d\rad g \dot{G} G
\int \frac{d^3{\bf k}}{(2\pi)^3} \int \frac{d^3{\bf k}'}{(2\pi)^{3}}
\nonumber \\
&&\times \delta_\delta^\lin({\bf k}-{\bf k}')\delta_{x_e}({\bf k'})
e^{i{\bf k}\cdot \hat{\bf n}\rad} \left[ \hat{\bf n} \cdot
\frac{\hat{\veck} - \hat{\veck'}}{|\veck - \veck'|}\right] \, , \nonumber \\
\label{eqn:rei}
\end{eqnarray}
where we have relate the velocity field to the density field using the continuity equation such that
\begin{eqnarray}
{\bf v} =  -i \dot G \delta(k,0){ {\bf k} \over k^2 }\,,
\label{eqn:continuity}
\end{eqnarray}
where the overdot represent the derivative with respect to radial distance $\rad$
and the linear density field is scaled as a function of time using the growth function $G(z)$, where
$\delta(k,r)=G(r)\delta(k,0)$ \cite{Pee80} 
\begin{equation}
G(r) \propto {H(r) \over H_0} \int_{z(r)}^\infty dz' (1+z') \left( {H_0
\over H(z')} \right)^3\,.
\end{equation}
Note that in the matter dominated epoch $G \propto a=(1+z)^{-1}$.

Similar to the Ostriker-Vishniac (OV) effect \cite{OstVis86}, this second order effect
related to reionization, called the patchy-reionization contribution,
avoids the strong cancellation associated with the linear Doppler effect.  
As discussed in Ref.~\cite{Santos:2003jb}, for high optical depth reionization histories, where the
reionization process is expected to be highly inhomogeneous,
this effect dominates secondary anisotropy fluctuations after the SZ contribution related to galaxy clusters;
it is higher than both the OV and Kinetic SZ contributions. We show these contributions in Fig~1(a).
When calculating the patchy reionization power spectrum, we model the free-electron distribution with
a power spectrum that involves the same halo-based bias as in Ref.~\cite{Santos:2003jb}; the
bias is dominated by halos with temperature at the level of $10^4$ K  and above,
where atomic cooling is expected and first objects form and subsequently reionize the universe. 
Since what enters in the power spectrum is
the bias factor times the growth of density perturbations, in this case the product $b_{X_e}(z)G(z)$
is in fact a constant as a function of redshift out to z of 20 or more
given the rareness of massive halos at high redshifts \cite{Oh:2003sa}.

We can expand the temperature perturbation due to reionization related scattering, $T^\dsz$, into
spherical harmonics given by $a_{lm} = \int d\bn T(\bn) \Ylmn {}^*(\bn)$ such that
\begin{eqnarray}
&&a_{lm}^\dsz = -i \int d\hat{\bf n}
\int d\rad\; (g\dot{G} G)
\int \frac{d^3{\bf k_1}}{(2\pi)^3}\int \frac{d^3{\bf
k_2}}{(2\pi)^3}\nonumber \\
&&\times \delta_\delta^\lin({\bf k_1})\delta_g({\bf k_2})
e^{i({\bf k_1+k_2})
\cdot \hat{\bf n}\rad} \left[ \frac{\hat{\bf n} \cdot \hat{\veck_1}}{k_1} \right]
Y_l^{m\ast}(\hat{\bf n}) \, ,
\end{eqnarray}
where we have symmetrizised by using $\veck_1$ and $\veck_2$
to represent $\veck-\veck'$ and $\veck'$ respectively.
Using
\begin{equation}
\hat{\bf n} \cdot \hat{\veck} = \sum_{m'} \frac{4\pi}{3} k
Y_1^{m'}(\hat{\bf n}) Y_1^{m'\ast}(\hat{\veck}) \, ,
\end{equation}
we can further simplify and rewrite the multipole moments as
\begin{eqnarray}
&&a_{lm}^\dsz = -i \frac{(4 \pi)^3}{3}
\int d\rad
\int \frac{d^3{\bf k}_1}{(2\pi)^3} \int \frac{d^3{\bf
k}_2}{(2\pi)^3}
\sum_{l_1 m_1}\sum_{l_2 m_2}\sum_{m'} \nonumber\\
&& \times
(i)^{l_1+l_2}
(g\dot{G}G)
\frac{j_{l_1}(k_1\rad)}{k_1}
j_{l_2}(k_2\rad)
\delta_\delta^\lin({\bf k_1})\delta_g({\bf k_2})
\nonumber\\
&& \times
Y_{l_1}^{m_1}(\hat{\veck}_1) Y_1^{m'}(\hat{\veck}_1)
Y_{l_2}^{m_2}(\hat{\veck}_2)  M_{l l_1 l_2 1}^{m^\ast m_1^\ast
m_2^\ast m'^{\ast}}(\hat{\bf n}) \, .
\label{eqn:almdsz}
\end{eqnarray}

\subsection{Cross Power Spectrum}

We can now construct the angular power spectra by considering the clustering aspects between
neutral Hydrogen in the 21 cm background and fluctuations in the secondary reionization related effects.
Under the  assumption that the two fields are statistically
isotropic, the cross-correlation is independent of $m$, and we can write the
the cross-power spectrum between 21 cm fluctuations and the secondary CMB
anisotropy fluctuations as
\begin{eqnarray}
\langle \alm{1}^{*, \dsz} \alm{2}^{\rm 21 cm}\rangle = \deld_{l_1 l_2} \deld_{m_1 m_2}
        C_{l_1}^{\dsz-21 cm}\, .
\end{eqnarray}

The cross-correlation can be written using
\begin{eqnarray}
&& \langle a^{\ast, \dsz}_{l_1m_1} a^{21 cm}_{l_2m_2} \rangle= \frac{(4 \pi)^5}{3}
\int d\rad_1 g \dot{G} G \int d\rad_2 W^{21 cm}(\rad)  \nonumber \\
&\times& \int \frac{d^3{\bf k_1}}{(2\pi)^3}\frac{d^3{\bf k_2}}{(2\pi)^3}
\frac{d^3{\bf k_1'}}{(2\pi)^3}\frac{d^3{\bf k_2'}}{(2\pi)^3}
\nonumber \\
&&\sum_{l_1'm_1' l_1'' m_1'' m_1''' l_2'm_2' l_2'' m_2''} \langle \delta_\delta^\lin({\bf k_1'})\delta_{x_e}({\bf k_2'})
\delta_\delta^{\ast \lin}({\bf k_1})\delta_{x_H}^\ast({\bf k_2})  \rangle
\nonumber \\
&\times& (-i)^{l_1'+l_1''} (i)^{l_2'+l_2''}
j_{l_2'}(k_1'\rad_2) \frac{j_{l_2''}(k_2'\rad_2)}{k_2'}
j_{l_1'}(k_1\rad_1)
j_{l_1''}(k_2\rad_1) \nonumber \\
&\times&
 Y_{l_2'}^{m_2'}(\hat{\veck_1'}) Y_{1}^{m_2'''}(\hat{\veck_2'})
Y_{l_2''}^{m_2''}(\hat{\veck_1'})
Y_{l_1'}^{m_1' \ast}(\hat{\veck_1})
Y_{l_1''}^{m_1'' \ast}(\hat{\veck_2}) \nonumber \\
&\times& M_{l_2 l_2' l_2'' 1}^{m_2 m_2'^{\ast} m_2''^{\ast} m_2'''^{\ast}}(\hat{\bf m}) M_{l_1 l_1' l_1''}^{m_1^\ast m_1' m_1''}
(\hat{\bf n}) \, .
\nonumber \\
\end{eqnarray}

We can separate out the contributions such that the total is made of
correlations
following $\langle v_g \delta_\delta\rangle \langle \delta_{x_e} \delta_{x_H} \rangle$
and $\langle v_g \delta_{x_H} \rangle \langle \delta_\delta \delta_{x_e} \rangle$
depending on whether we consider cumulants by combining $\veck_1$ with $\veck_1'$
or $\veck_2'$ respectively. After some straightforward but tedious
algebra, and noting the orthonormality of the Wigner-3j symbols
\begin{equation}
\sum_{m_1' m_2'}
\left(
\begin{array}{ccc}
l_1' & l_2' & l_1 \\
m_1' & m_2'  &  m_1
\end{array}
\right)
\left(
\begin{array}{ccc}
l_1' & l_2' & l_2 \\
m_1' & m_2'  &  m_2
\end{array}
\right)
= \frac{\delta_{m_1 m_2} \delta_{l_1 l_2}}{2l_1+1} \, ,
\label{eqn:ortho}
\end{equation}
we can write
\begin{eqnarray}
&&C_l^{\rm 21cm-\dsz} = \frac{2^2}{\pi^2} \sum_{l_1 l_2}
\left[\frac{(2l_1+1)(2l_2+1)}{4\pi}\right]
\left(
\begin{array}{ccc}
l & l_1 & l_2 \\
0 & 0  &  0
\end{array}
\right)^2 \nonumber \\
&\times& \int d\rad_1 g \dot{G} G
\int d\rad_2 W^{21 cm}(\rad)
\int k_1^2 dk_1 \int k_2^2 dk_2 \nonumber \\
&\times& P_{\delta\delta}^\lin(k_1)
P_{x_e x_H}(k_2)j_{l_1}(k_2\rad_2) j_{l_1}(k_2\rad_1)
\frac{j_{l_2}'(k_1\rad_1)}{k_1} j_{l_2}(k_1\rad_2) \, . \nonumber \\
\end{eqnarray}
Here, we have only considered the contribution related to the $\langle v_g \delta
\rangle \langle x_e  x_H \rangle$ term; we ignore the contribution related to
the  $\langle v_g x_H \rangle \langle \delta x_e\rangle$ term due to the mismatch in fluctuation
scales between velocity fluctuations (that trace large angular linear scales)  and the distribution of neutral
Hydrogen  (which is expected to be patchy and concentrated to smaller halos); By ignoring this term, we have
underestimated the cross-power by, at most, a factor of 2 and this difference is unlikely to change our conclusions regarding the
detectability. In simplifying the integrals involving spherical harmonics, we have
made use of properties of Clebsh-Gordon coefficients, in
particular, those involving $l=1$. The integral involves two distances
and two Fourier modes and is summed over the Wigner-3$j$ symbol to
obtain the observable cross-power spectrum.

Similar to the Limber approximation \cite{Lim54},
in order to simplify the calculation, we use an equation
involving completeness of spherical Bessel functions:
\begin{equation}
\int dk k^2 F(k) j_l(kr) j_l(kr')  \approx {\pi \over 2} \da^{-2}
\deld(r-r')
                                                F(k)\big|_{k={l\over
d_A}}\,,
\end{equation}
where the assumption is that $F(k)$ is a slowly-varying function.
Applying this to the integral over $k_2$ gives
\begin{eqnarray}
&&C_l^{\rm 21cm-\dsz} = \frac{2}{\pi} \sum_{l_1 l_2}
\left[\frac{(2l_1+1)(2l_2+1)}{4\pi}\right]
\left(
\begin{array}{ccc}
l & l_1 & l_2 \\
0 & 0  &  0
\end{array}
\right)^2 \nonumber \\
&\times& \int d\rad_1 \frac{(g G \dot{G})W^{21 cm}}{d_A^2}  
P_{x_e x_H}\left[ \frac{l_1}{d_A}; \rad_1 \right] J_{l_2}(\rad) 
\label{eqn:redallsky}
\end{eqnarray}
where the mode-coupling integral is
\begin{eqnarray}
J_{l_2}(\rad) &=& \int k_1^2 dk_1 P_{\delta\delta}^\lin(k_1) 
\frac{j_{l_2}'(k_1\rad_1)}{k_1}j_{l_2}(k_1\rad_1) \, . \nonumber \\
\label{eqn:Jl}
\end{eqnarray}

\begin{figure}[!h]
\centerline{\psfig{file=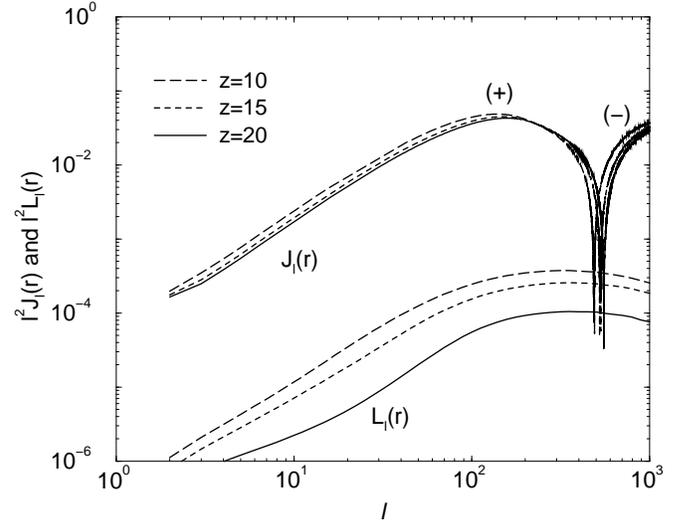,width=3.4in,angle=-90}}
\caption{The integrals $J_{l}(\rad)$, top lines (Eq.~\ref{eqn:Jl}), and $L_{l}(\rad)$, bottom lines (Eq.~\ref{eqn:Ll}),
as a function of the multipole when $\rad$ is given by the redshift indicated.  The mode coupling integral related to the
cross-power spectrum, $J_{l}(\rad)$, oscillate between positive and negative values and we label 
 positive/negative parts with (+)/(-). These functions captures the cross-correlation between
velocity field related to the CMB secondary anisotropies and either the density field fluctuations, 
in the case of the cross-power spectrum, or with the velocity field traced by the large scale Doppler contribution,
in the case of the bispectrum.}
\label{fig:modecouple}
\end{figure}

\subsection{Bispectrum}

As we will soon discuss, the cross power spectrum is not an ideal probe of the cross-correlation between fluctuations in the
ionized and neutral regions around the era of reionization due to the geometric cancellation associated with large-scale
velocity projection. Instead, we consider the possibility for a measurement of the 
three-point correlation function or, in Fourier space, the bispectrum that is constructed in a way such that it avoids the cancellation
related to the cross power spectrum.
For reference,  the three-point correlation function is related to multipole moments via
\begin{eqnarray}
B(\bn,\bm,\bl) &\equiv& \langle T(\bn)T(\bm)T(\bl) \rangle \\
               &\equiv&
                \sum 
                \langle \alm{1} \alm{2} \alm{3} \rangle
                \Ylm{1}(\bn) \Ylm{2}(\bm)  \Ylm{3}(\bl)\,,\nonumber
\end{eqnarray}
where the sum is over $(l_1,m_1),(l_2,m_2),(l_3,m_3)$.
Statistical isotropy again allows us to express the correlation in terms an $m$-independent function,
\begin{eqnarray}
\langle \alm{1} \alm{2} \alm{3} \rangle  = \wjm \bi\,,
\end{eqnarray}
where the quantity $\bi$ is described as the angular averaged bispectrum \cite{Cooray:1999kg}
and we have made use of the orthonormality relation of Wigner-3j symbols (Eq.~\ref{eqn:ortho}).
As discussed in Ref.~\cite{Cooray:1999kg}. the angular bispectrum, $\bi$, contains all the information available
in the three-point correlation function and frequently used quantities such as the skewness and the collapsed three-point function
can be written in terms of the bispectrum (such as a filtered sum of certain bispectrum configurations) \cite{Cooray:2000uu}.

We consider a bispectrum of the form Doppler-Reionization-21 cm fluctuations. The bispectrum can be written as
\begin{eqnarray}
&&a^{\rm 21cm}_{l_1 m_1}a^{\dop}_{l_2 m_2}a^{\rm Rei}_{l_3 m_3} = \nonumber \\
&& {4 \over \pi^2} \int k_1^2 dk_1 \int k_2^2 dk_2 P_{\delta \delta}(k_1) P_{x_H x_e}(k_2)
I_{l_1}^{\rm 21 cm}(k_1)
\nonumber \\
&&\times
I_{l_2}^{\dop}(k_2) I_{l_1,l_2}^{\rm Rei}(k_1,k_2) M_{l_1 l_2 l_3}^{m_1 m_2 m_3}
\label{eqn:ovtriplet}
\end{eqnarray}
where
\begin{eqnarray}
I^{\rm Rei}_{l_1,l_2}(k_1,k_2) &=& \int d\rad W^{\rm Rei} j_{l_2}(k_2\rad)
j'_{l_1}(k_1\rad) \nonumber\,,\\
W^{\rm Rei}(k_1,r) &=& -{1 \over k_1} g\dot{G}G \,,
\end{eqnarray}
and, similarly, following multipole moments of Eq.~\ref{eqn:doppler},
\begin{eqnarray}
I^{\dop}_{l_1}(k_1) &=& \int d\rad W^{\rm Dop} j'_{l_1}(k_1\rad) \nonumber\,,\\
W^{\dop}(k_1,r) &=& -{1 \over k_1} g\dot{G} \,.
\end{eqnarray}
In simplifying the integrals involving spherical harmonics, note that we have
again made use of the properties of Clebsch-Gordon coefficients, in
particular, those involving $l=1$. 

The bispectrum is given by
\begin{eqnarray}
B_{l_1 l_2 l_3} &=& \sum_{m_1 m_2 m_3} \wjm 
\left< a^{\rm 21 cm}_{l_1 m_1}a^{\dop}_{l_2 m_2}a^{\rm Rei}_{l_3 m_3}  \right> \nonumber \\
&=& \sqrt{\frac{(2l_1 +1)(2 l_2+1)(2l_3+1)}{4 \pi}}
\left(
\begin{array}{ccc}
l_1 & l_2 & l_3 \\
0 & 0  &  0
\end{array}
\right) b_{l_1,l_2}\, . \nonumber \\
\label{eqn:ovbidefn}
\end{eqnarray}
Here,
\begin{eqnarray}
 b_{l_1,l_2}
&=& \frac{4}{\pi^2} \int k_1^2 dk_1 \int k_2^2 dk_2 P_{\delta \delta}(k_1) P_{x_H x_e}(k_2) \nonumber \\
&&\times I^{\rm Rei}_{l_1,l_2}(k_1,k_2) I_{l_1}^{\rm 21cm}(k_1) I_{l_2}^{\dop}(k_2) \, .
\label{eqn:finalintegral}
\end{eqnarray}

As with the cross-power spectrum calculation, we use
Limber approximation to describe the coupling between 21 cm and reionization ionized and neutral fractions.
The bispectrum term is then,
\begin{eqnarray}
b_{l_1,l_2}&= & \frac{2}{\pi} \int dr \frac{(g G \dot{G})W^{\rm 21cm}}{\da^2} P_{x_e x_H} \left[\frac{l_1}{\da};\rad\right] L_{l_2}(\rad)\nonumber \\
\end{eqnarray}
where the new mode coupling integral is now
\begin{equation}
L_{l_1}(\rad)=\int k^2 dk P_{\delta \delta}(k) \frac{j'_{l_1}(k\rad)}{k} \int d\rad_1 
(g\dot{G})(\rad_1) \frac{j'_{l_1}(k\rad_1)}{k}  \, .
\label{eqn:Ll}
\end{equation}
This can be directly compared to the mode coupling integral associated with the cross-power spectrum.

\begin{figure}[!h]
\centerline{\psfig{file=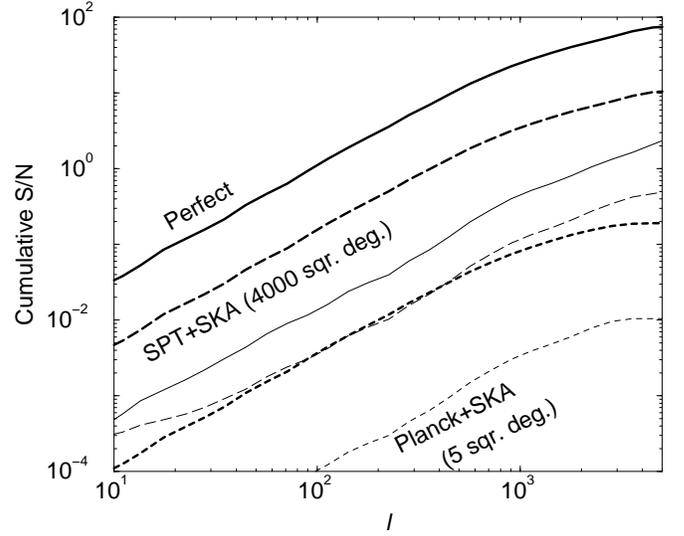,width=3.4in,angle=-90}}
\caption{The cumulative signal-to-noise ratio as a function of the multipole $l$. With thin lines, we show the case for the
angular cross power spectrum between 21 cm fluctuations and small-scale CMB anisotropies while with thick lines, we consider
the case involving the bispectrum between 21 cm fluctuations, small-scale CMB, and large-scale CMB (in this case the multipole
shown in the x-axis is that of $l_3$ with the two other multipoles summed over). In both cases, the top lines are for
a case involving all-sky observations with no instrumental noise, but limited by the cosmic variance. The middle long-dashed lines
assume instrumental noise level of the South Pole Telescope (SPT), shown in Fig.~1(a), combined with SKA noise for over 4000 sqr. degrees.
The bottom lines show the case involving Planck data when combined with a map of the SKA 5 sqr. degree field of view. The signal-to-noise
ratio for the cross power spectrum is smaller than that of the bispectrum.}
\label{fig:sn}
\end{figure}

\section{Results}
\label{sec:results}

In Fig.~1, we show power spectra of CMB temperature anisotropies (top panel) and brightness temperature fluctuation spectra 
at 140 MHz corresponding to redshifted 21 cm line emission from a redshift $\sim$ 9 (bottom panel). When calculating
21 cm fluctuation spectrum, we assume a bandwidth for observations of 1 MHz centered around 140 MHz; this roughly corresponds to
a redshift interval of 0.06. For reference, in Fig~1(b), we also show the noise level related to a single month of observation with the planned
Square Kilometer Array (SKA); We take the approach of Ref.~\cite{Zaldarriaga:2003du} and assume a 
system temperature of $T_{\rm sys}=200$ K. 

In Fig.~2, we illustrate the mode coupling integrals related to both the cross power spectrum and bispectrum between 21 cm
fluctuations and CMB anisotropies related to reionization. In the case of the cross power spectrum, the mode coupling term
(Eq.~31) involves an integral over the product of $j_{l}'$ and $j_l$ terms. The $j_l'$ term here is related to the line of
sight projection of velocity fluctuations from the CMB side while the $j_l$ term is associated with the line of
sight projection of density fluctuations in the 21 cm map. The two projections, unfortunately, are not well
correlated (analytically, this can be understood based on the phase shift between $j_{l}'$ and $j_l$ terms, while, physically,
the density inhomogeneities peak at a smaller physical scale than the velocity field). The resulting effect is to 
produce a mode coupling integral that oscillate between positive and negative values. Since the final cross power spectrum is
a weighted summation over this mode coupling term, this results in a net cancellation and a reduction of the cross correlation.
Instead of the all-sky derivation used here, the same calculation can be considered under a flat-sky approximation.
In this case, the correlation term involves a term that scales as $\int d^3{\bf k} \int d\bn \langle [\bn \cdot \vec{v}] \delta (k) \rangle$,
which simplifies at the small angle limit to 
a zero when one integrates over the $\cos(\theta)$ angle between the line of sight and the velocity field.

The relative cancellation between velocity and density fluctuations can be avoided if one introduces an additional probe of the
velocity field to the correlation measurement. We have suggested the bispectrum where, in addition to 21 cm fluctuations and small
scale CMB anisotropies related to reionization, one makes use of large angular scale CMB anisotropies related to the Doppler effect.
The mode coupling integral now involves two $j_{l}'$ terms that capture the velocity field from both the large scale Doppler
effect and small scale anisotropies which are due to a modulation of the same velocity field. The mode coupling integral no
longer oscillates and avoids the cancellation related to the cross power spectrum. While this is an advantage, there is a disadvantage
related to this bispectrum since it involves a map of large scale temperature fluctuations in addition  to small scale anisotropies.
In terms of the detection, the signal-to-noise ratio is limited by the cosmic variance related to the large scale Doppler
effect which can be understood mainly as a confusion between these fluctuations and the ones generated at degree angular scales
at the surface of last scattering such as the acoustic peak structure.

In order to consider the extent to which the cross-correlation can be detected and studied to understand the reionization process,
we estimate the signal-to-noise ratio for a detection of both the cross power spectrum and the bispectrum.
In the case of the power spectrum, the signal-to-noise ratio is
\begin{equation}
\left(\frac{{\rm S}}{{\rm N}}\right)^2 = \sum_{l} (2l+1) f_{\rm sky} \frac{\left(C_l^{\rm 21cm-Rei}\right)^2}{
  \left(C_l^{\rm 21cm-Rei}\right)^2+C_l^{\rm CMB,tot}C_l^{\rm 21cm,tot}} \, ,
\end{equation}
while the signal-to-noise ratio for the bispectrum is calculated as
\begin{equation}
\left(\frac{{\rm S}}{{\rm N}}\right)^2 = f_{\rm sky} \sum_{l_1 l_2 l_3}
        \frac{\left(B_{l_1 l_2 l_3}\right)^2}{
          C_{l_1}^{\rm 21cm,tot}
          C_{l_2}^{\rm CMB,tot}
          C_{l_3}^{\rm CMB,tot}}\,,
\label{eqn:chisq}
\end{equation}
Here, $C_l^{i,tot}$ represents all contributions to the power spectrum of the $i$th field,
\begin{eqnarray}
C_l^{i,tot} = C_l^i + C_l^{\rm noise} + C_l^{\rm foreg}\, ,
\label{eqn:cltot}
\end{eqnarray}
where $C_l^{\rm noise}$ is the noise contribution and $C_l^{\rm foreg}$ is the confusing foreground contribution,
and $f_{\rm sky}$ is the fraction of sky covered by the data. 
In each of the cases involving CMB or 21 cm maps, we make use of the noise and foreground spectra shown in Fig.~1;
In the case of CMB foregrounds, we make the assumption that SZ contribution can be perfectly removed based on multifrequency maps
(such as with Planck). This is important as this leads to an improvement in the signal-to-noise ratio at arcminute angular scales with
a reduction in an additional confusion. 

The signal-to-noise ratios are summarized in Fig.~3 for both the cross power spectrum and the bispectrum. Here, in addition to
realistic scenarios, we also consider the maximal signal-to-noise ratio one could achieve by using no instrumental noise all-sky
maps; In the case of the cross power spectrum, the maximum signal-to-noise ratio is order a few while, for the bispectrum, this
is at a level slightly below 100. In addition to the partial reduction of the geometric cancellation associated with the
mode coupling integral, the signal-to-noise ratio with the bispectrum is higher as one has more modes from which to construct and
measure the bispectrum when compared to that of the power spectrum. 

Using Planck and SPT maps, combined with SKA 21 cm maps,
we also show the level of the expected signal-to-noise ratios  in Fig.~3. The bispectrum constructed with the 4000 sqr. degree
map of SPT, combined with Planck large scale CMB map, could provide the first opportunity to detect the suggested cross-correlation. 
We note here that techniques now exist to construct the bispectrum reliably from CMB and large scale
structure maps and they have been successfully applied to understand the presence of non-Gaussianities  in current data.
Thus, we do not consider the measurement of the proposed bispectrum, involving CMB temperature (both at large and small angular scales)
and the 21 cm background to be any more complicated than what is already achieved \cite{Komatsu:2001wu}.
While the measurement is not complicated,  
unfortunately, since the cumulative signal-to-noise ratio is at the level of ten, it is unlikely that a detection of the bispectrum can
be used to establish in detail the cross spatial power spectrum of neutral and ionized regions during reionization. 

While the signal-to-noise is not high, we still recommend the proposed measurement in future data as we may have underestimated the
cross-correlation, and thus the expected signal-to-noise ratio,
 by a variety of effects. For example, if there is a significant contribution to CMB anisotropies
from reionization, beyond the patchy reionization contribution we have suggested, such as in the SZ map, then this would
be evident in the cross-correlation. The cross-correlation can also be used as a mechanism to separate out small scale CMB anisotropies
related to, for example, lensing effect on CMB from that of reionization; this separation is generally hard to achieve since
both effects have the same thermal spectrum and peak at similar angular scales.

\section{Summary}
\label{sec:summary}
During the transition from a neutral to a fully reionized universe, scattering of
cosmic microwave background (CMB) photons via free-electrons lead to a new anisotropy contribution to the temperature distribution.
If the reionization process is inhomogeneous and patchy, the era of reionization is also visible via
brightness temperature fluctuations in the redshifted 21 cm line emission
of neutral Hydrogen. Since regions containing electrons and neutral Hydrogen are
expected to trace the same underlying density field, the two are (anti) correlated and this is expected to be
reflected in the anisotropy maps in terms of a cross-correlation between arcminute-scale CMB  temperature and the
21 cm background. 

In terms of the angular cross-power spectrum, unfortunately,
this correlation is insignificant due to a geometric cancellation associated with second order CMB anisotropies.
Thus, it is unlikely that the cross-correlation spectrum between small-scale CMB and 21 cm fluctuations will be
measurable even with maps involving a high signal-to-noise per pixel.
The same cross-correlation between ionized and neutral regions, however, can  be studied using a bispectrum
involving large scale velocity field of ionized regions from the Doppler effect,
arcminute scale CMB anisotropies involving reionization signal, and the 21 cm background.
While the geometric cancellation is partly avoided, the signal-to-noise ratio related to this bispectrum is reduced via large
cosmic variance related to the velocity fluctuations traced by the Doppler effect.
Unless velocity fluctuations can be independently established,
it is unlikely that the correlation information related to the relative distribution of ionized electrons and regions containing
neutral Hydrogen can be obtained with a combined study involving CMB and 21 cm fluctuations.

\acknowledgments
This work was supported in part by DoE DE-FG03-92-ER40701 and 
a senior research fellowship from the Sherman Fairchild foundation.

\end{document}